\documentclass[a4paper,11pt]{article}
\pdfoutput=1 

\usepackage[utf8]{inputenc}
\usepackage{jinstpub} 
\usepackage{subcaption}
\usepackage[table,xcdraw]{xcolor}
\usepackage{lineno}
\usepackage{multirow}

\newcommand{\nucl}[3]{
$\!^{#1}_{#2}{\rm #3}$}

\title{The Impact of Helium Exposure on the PMTs of the SuperNEMO Experiment}

\author[a, b]{X.~Aguerre,}
\author[c]{A.S.~Barabash}
\author[d]{A.~Basharina-Freshville,}
\author[e]{M.~Bongrand,}
\author[e]{Ch.~Bourgeois,}
\author[e]{D.~Breton,}
\author[f]{R.~Breier,}
\author[g]{J.~Busto,}
\author[a]{C.~Cerna,}
\author[d]{M.~Ceschia,} 
\author[a]{E.~Chauveau,}
\author[d]{A.~Chopra,}
\author[d]{L.~Dawson,}
\author[h]{D.~Duchesneau,}
\author[i]{J.J.~Evans,}
\author[c]{D.~Filosofov,}
\author[e]{X.~Garrido,}
\author[e]{C.~Girard-Carillo,}
\author[a]{M.~Granjon,}
\author[e]{M.~Hoballah,}
\author[k]{R.~Hodák,}
\author[d]{G.~Horner,}
\author[d]{M.H.~Hussain,}
\author[d]{A.~Islam,}
\author[h]{A.~Jérémie,}
\author[e]{S.~Jullian,}
\author[f]{J.~Kaizer,} 
\author[c]{A.~Klimenko,} 
\author[c]{O.~Kochetov,}
\author[k, l]{F.~Koňařík,} 
\author[c]{S.I.~Konovalov,}
\author[k, l]{T.~Křižák,} 
\author[c]{S.~Kovalenko,}
\author[a]{A.~Lahaie,}
\author[m]{K.~Lang,}
\author[j]{Y.~Lemière,}
\author[b]{P.~Li,}
\author[e]{J.~Maalmi,}
\author[k]{M.~Macko,}
\author[k]{F.~Mamedov,}
\author[a]{C.~Marquet,}
\author[j]{F.~Mauger,}
\author[k, n]{A.~Mendl,} 
\author[o]{B.~Morgan,}
\author[c]{I.~Nemchenok,}
\author[p]{M.~Nomachi,}
\author[k]{V.~Palu\v{s}ov\'{a},}
\author[b]{C.~Patrick,}
\author[d]{T.~Pavicic,}
\author[a]{F.~Perrot,}
\author[f, k]{M.~Petro,}
\author[a]{F.~Piquemal,}
\author[f]{P.~Povinec,}
\author[b]{S.~Pratt,}
\author[m]{M.~Proga,}
\author[d]{W.S.~Quinn,}
\author[o]{Y.A.~Ramachers,}
\author[q]{C.L.~Riddle,}
\author[c]{N.~Rukhadze,} 
\author[d]{R.~Saakyan,}
\author[m]{R.~Salazar,}
\author[r]{J.~Sedgbeer,}
\author[k]{Yu.~Shitov,}
\author[e]{L.~Simard,}
\author[f, k]{F.~\v{S}imkovic,}
\author[c]{A.~Smolnikov,}
\author[i, r]{S.~Söldner-Rembold,}
\author[k]{I.~\v{S}tekl,}
\author[s]{J.~Suhonen,}
\author[g]{H.~Tedjditi,}
\author[d]{J.~Thomas,}
\author[c]{V.~Timkin,}
\author[c]{V.I.~Tretyak,}
\author[t, u]{Vl.I.~Tretyak,}
\author[b]{G.~Turnbull,}
\author[c]{V.I.~Umatov,}
\author[c]{Y.~Vereshchaka,}
\author[d]{D.~Waters}

\affiliation[a]{Université de Bordeaux, CNRS/IN2P3, LP2i Bordeaux, UMR 5797, F-33170 Gradignan, France}
\affiliation[b]{University of Edinburgh, Edinburgh, EH9 3FD, United Kingdom}

\affiliation[c]{Participant in the NEMO-3/SuperNEMO collaboration}
\affiliation[d]{University College London, London, WC1E 6BT, United Kingdom}
\affiliation[e]{Université Paris-Saclay, CNRS, IJCLab, F-91405 Orsay, France}
\affiliation[f]{Faculty of Mathematics, Physics and Informatics, Comenius University, SK-842 48 Bratislava, Slovakia}
\affiliation[g]{Aix-Marseille Université, CNRS, CPPM, F-13288 Marseille, France}
\affiliation[h]{Université de Savoie, CNRS/IN2P3, LAPP, UMR 5814, F-74941 Annecy-le-Vieux, France}
\affiliation[i]{University of Manchester, Manchester, M13 9PL, United Kingdom}
\affiliation[j]{Université de Caen Normandie, ENSICAEN, CNRS/IN2P3, LPC Caen UMR6534, F-14000 Caen, France}
\affiliation[k]{Institute of Experimental and Applied Physics, Czech Technical University in Prague, CZ-11000 Prague, Czech Republic}
\affiliation[l]{Faculty of Nuclear Sciences and Physical Engineering, Czech Technical University in Prague, Brehova, 7 115,  Czech Republic}
\affiliation[m]{University of Texas at Austin, Austin, TX 78712, U.S.A.}
\affiliation[n]{Faculty of Mathematics and Physics, Charles University, CZ-12116, Prague, Czech Republic}
\affiliation[o]{University of Warwick, Coventry, CV4 7AL, United Kingdom}
\affiliation[p]{Osaka University, 1-1 Machikaneyama Toyonaka, Osaka 560-0043, Japan}
\affiliation[q]{Idaho National Laboratory, Idaho Falls, ID 83415, U.S.A.}
\affiliation[r]{Imperial College London, London, SW7 2AZ, United Kingdom}
\affiliation[s]{Jyväskylä University, FIN-40351 Jyväskylä, Finland}
\affiliation[t]{Institute for Nuclear Research of NASU, 03028 Kyiv, Ukraine}
\affiliation[u]{INFN - Laboratori Nazionali del Gran Sasso, 67100 Assergi (AQ), Italy}

\collaboration{SuperNEMO Collaboration:}

\emailAdd{william.quinn.14@ucl.ac.uk}

\abstract{The performance of Hamamatsu 8" photomultiplier tubes (PMTs) of the type used in the SuperNEMO neutrinoless double-beta decay experiment (R5912-MOD), is investigated as a function of exposure to helium (He) gas. Two PMTs were monitored for over a year, one exposed to varying concentrations of He, and the other kept in standard atmospheric conditions as a control. Both PMTs were exposed to light signals generated by a \nucl{207}{}{Bi} radioactive source that provided consistent large input PMT signals similar to those that are typical of the SuperNEMO experiment. The energy resolution of PMT signals corresponding to 1~MeV energy scale determined from the \nucl{207}{}{Bi} decay spectrum, shows a negligible degradation with He exposure; however the rate of after-pulsing shows a clear increase with He exposure, which is modelled and compared to diffusion theory. A method for reconstructing the partial pressure of He within the PMT and a method for determining the He breakdown point, are introduced. The implications for long-term SuperNEMO operations are briefly discussed.}


\begin{document}
\maketitle
\flushbottom

\section{Introduction}
    The effect of helium (He) poisoning on photomultiplier tube (PMT) performance is well established~\cite{Incandela:1987dh, ospanov2019studiesheliumpoisoninghamamatsu, Ma:2009aw}. The performance of a PMT is dependent on the quality of the vacuum within the tube, amongst other attributes. Any gases, either relics from the manufacturing process or those that permeate through the PMT structure, can interfere with the photoelectrons accelerated between the photocathode and the dynode stages. The ions drift in the electric field of the PMT and can release secondary electrons upon impact with photocathode or electrode surfaces. In sufficiently large quantities, these ions will produce a continuous Townsend discharge which can render the PMT unusable. In smaller quantities, the primary PMT signals will be accompanied by secondary, and perhaps tertiary, pulses. These after-pulses are a cause of performance degradation, affecting pulse shapes, pulse amplitudes, pulse-time identification and other features that are extracted from the recorded PMT pulses.    
    \par
    The SuperNEMO Demonstrator, located in the Laboratoire Souterrain de Modane (LSM) underground laboratory, is a proof-of-concept module designed to search for the lepton-number-violating process of neutrinoless double-beta decay. The detector comprises a multi-wire drift chamber surrounding the  \nucl{82}{}{Se} source foils, which in turn is surrounded by a plastic-scintillator based calorimeter~\cite{SuperNEMO:2010wnd}. The ability to reconstruct long trajectories of electron tracks gives SuperNEMO, uniquely among experiments in the field, the ability to fully reconstruct the double-beta event kinematics. However, to accurately reconstruct MeV-scale electrons demands a low-$Z$ ionisation medium, and the standard SuperNEMO tracker-gas mixture comprises $95\%$ He by volume. The calorimeter consists of arrays of PMTs each optically coupled to a polystyrene-based scintillator block to form what is referred to as an Optical Module (OM). 440 of these OMs, making up the majority of the SuperNEMO's largest calorimeter walls, contain 8" Hamamatsu PMTs (model number R5912-MOD) with the final design resulting from an extensive R\&D programme~\cite{Barabash:2017sxf}. Although the tracker volume is separated from the space occupied by the PMTs, leaks and diffusion of He through sealing interfaces can give rise to enhanced He concentrations in the vicinity of the calorimeter PMTs. It is therefore imperative to have a quantitative understanding of the potential impact of He exposure on the performance of the SuperNEMO PMTs.
    \par
    This article reports on the results from a test-stand in which two SuperNEMO model PMTs were monitored over the course of a year. A \nucl{207}{}{Bi} source illuminated small scintillator blocks bonded to the PMT bulbs. These generated large primary light pulses with a number of photoelectrons similar to those that are measured in-situ in SuperNEMO. The \nucl{207}{}{Bi} decay spectrum was reconstructed in order to assess the impact on energy resolution, and Matched Filtering (MF) techniques were used to analyse the rate and distribution of after-pulses that was then compared to a simple model. This study is one of the first of its kind to analyse the impact of He exposure on large primary pulses in PMTs of this type.
    \par
    In section~\ref{sec:helium_pmt} the underlying theory of He permeation is discussed and the model used to produce the resulting data is presented. Furthermore, the volume of He inside the PMT is estimated by comparing the electron-He ionisation cross-section and the measured after-pulse rate. In sections~\ref{sec:experiment} and \ref{sec:ap_rates}, the experimental setup is described and the results of the data analysis, including the extraction of He permeation model parameters, are presented.

\section{The Impact of Helium on Photomultiplier Tubes}
\label{sec:helium_pmt}
    When photoelectrons are liberated from the PMT photocathode, they are accelerated towards the dynode stages, where they undergo cascade multiplication before being measured as an electrical pulse at the anode. Since PMT voltages are far in excess of typical ionisation energies, these electrons can cause impact ionisation of any residual gas atoms and molecules inside the PMT vacuum. The positive ions that are formed drift towards the photocathode where they can liberate additional electrons which are accelerated and multiplied in the same way as the primary photoelectrons. Due to the positive ion drift times, the experimental signature of this phenomenon is a secondary pulse, or an {\em after-pulse}, which is adjacent to but well separated in time from the primary pulse. Secondary photoelectrons that comprise after-pulses can themselves cause further ionisation of gas inside the PMT; with sufficiently high partial pressures this process becomes a self-sustaining Townsend discharge which can render the PMT unusable \cite{Townsend}. 
    
    \subsection{Electron-He Ionisation Cross Sections}
        \label{subsec:cross-sec}
        The generation of He ions in a PMT event is a random process following a Poisson distribution. The expected number of ions per event, $N_{\text{He}^+}$, for a given number of primary photoelectrons, $n_{\text{p.e}}$, is given by:
        \begin{equation}
            \label{equ:N_int_x}
            N_{\text{He}^+} = n_{\rm p.e.}\frac{p_i}{kT}\int_{x_0}^{d}\sigma(x) dx = I_{\sigma}n_{\rm p.e.}p_i\; ,
        \end{equation}
        where $x$ is the variable distance of the He ion between $x_0$ and $d$, where $x$ is the position within the electric field of the PMT where the electrons have the minimum energy to ionise a He atom, and $d$ is the full distance from photocathode to the first dynode stage. Additionally, $k$ is Boltzman's constant, $\sigma$ is the energy-dependent ionisation cross-section, $I_{\sigma}$ is the integrated cross-section corrected for the temperature ($T$) and $p_i$ is the internal partial pressure of He within the PMT glass volume.
        \par
\begin{figure}
            \centering
            \includegraphics{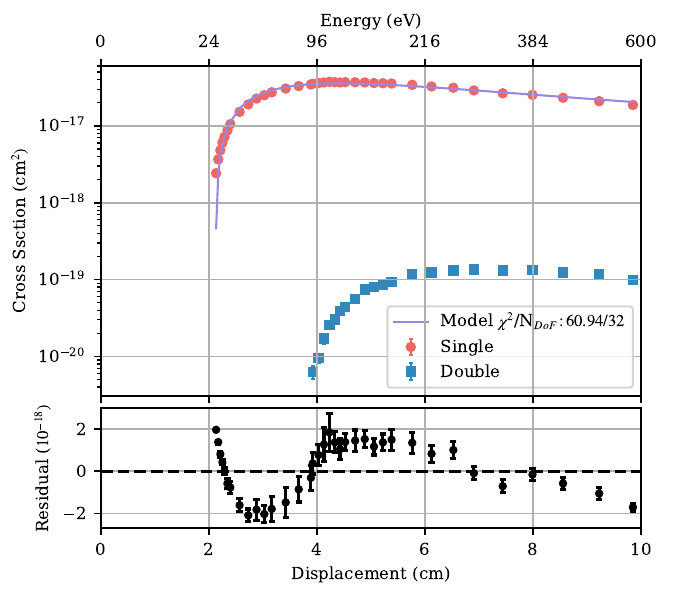}
            \caption{Electron-He ionisation cross-section data taken from \cite{Shah_1988} and fit with a functional form described in the text. The onset of single ionisation processes at energies corresponding to the $1^{\rm st}$ ionisation energies of He is clearly visible. The cross-section for double ionisation is also shown; since this is at least two orders of magnitude smaller, is neglected in this study.}
            \label{fig:xsec}
        \end{figure}

 
        The ionisation cross-section is a function of electron energy, which corresponds to a function of distance travelled, $x$, according to a simple hemispherical electrostatic model of the PMT~\cite{Incandela:1987dh}:
        \begin{equation}
            E(x) = V_0\bigg(\frac{x}{d}\bigg)^2 [{\rm {eV}}],
        \end{equation}
        where $E(x)$ is the electric field strength along the radial axis, $x$, $V_0$ is the potential difference between the photocathode and the first dynode. The energy-dependent cross section is determined from empirical data taken from~\cite{Shah_1988}. This data was fitted to the functional form also from~\cite{Shah_1988}: 
        \begin{equation}
            \sigma(E(x)) \approx \frac{A\ln\big(C[E - B]\big)}{E - B} = \frac{A\ln\big(C[V_0\big(\frac{x}{d}\big)^2 - B]\big)}{V_0\big(\frac{x}{d}\big)^2 - B},
        \end{equation}
        where $A$, $B$ and $C$ are model shaping parameters whose best-fit values are $(4.9\pm0.1)\times10^{-15}~\text{cm}^2~\text{eV}$, $(-25\pm1)~\text{eV}$ and $(51\pm1)~\text{eV}^{-1}$ respectively. The data and fit are shown in figure~\ref{fig:xsec}. 
        The integral in equation~\ref{equ:N_int_x} can then be performed numerically, starting from the displacement $x_{0}$, where the electrons have a kinetic energy equal to $24.587$~${\rm eV}$, the first ionisation energy of He, and integrating up to the first-dynode stage displacement/potential. For this study, the distance between the photocathode and the 1st dynode stage, $d$, is estimated to be 10~cm with an expected variation of at most 1~cm due to the imperfect hemispherical assumption of the PMT glass, and the first dynode stage voltage, $V_0$, is approximately $600$~${\rm V}$ in (see section~\ref{sec:SuperNEMO_PMTs}). Performing the cross section integral, a value of $(2.17~\pm~0.03_{\text{stat}}~\pm~0.10_{\text{sys}})^{-16}\;\text{cm}^{3}$ is obtained, where the statistical and systematic uncertainties are determined from the fit output and the uncertainty in the value of $d$, respectively. For a room temperature of 293 K, a value of $I_{\sigma}~=~(0.054~\pm~0.001)\;\text{Pa}^{-1}$ is obtained.
    
    \subsection{Internal Pressure Limit}
    \label{subsec:int_p_limit}
        The internal partial pressure limit for a PMT to breakdown depends on the geometry of the PMT and is in general not precisely known. ET Enterprises suggests a value of (0.001-0.01) Torr, or (0.133-1.33 Pa), as the point where significant after-pulses are present and (0.01-0.1 Torr), or (1.33-13.3 Pa), where the PMT is inoperable \cite{ETEnterprises}. A methodology to determine the breakdown pressure is constructed in this section.
        \par
        As He ions approach the photocathode, their proximity draws out a number of secondary electrons, $n_{\text{s.e}}$, due to the photocathode's low work function. The number of electrons released will depend on the Helium ion energy; we parameterise an average number, $\lambda = n_{\text{s.e}}/N_{\text{He}^+}$, which is unknown a-priori and estimated from our data in section \ref{sec:lamda}. These secondary electrons behave identically to the primary photoelectrons in the PMT event and can ionise further He atoms while accelerating to the first dynode stage. This begins a chain of successive ionisation and electron releases. This recursive behaviour will produce a total number of He ions, $N^{\text{He}^+}_T$, associated to a single PMT event, which can be written as:
        \begin{equation}
            N^{\text{He}^+}_T = I_{\sigma}n_{\text{p.e}}p_i\sum_{m = 0}^{\infty}(I_{\sigma}\lambda p_i)^m \, ,
        \end{equation}
        This geometric series has a finite limit for $N^{\text{He}^+}_T$ given by:
        \begin{equation}
            N^{\text{He}^+}_T = \frac{I_{\sigma}n_{\text{p.e}}p_i}{1 - I_{\sigma}\lambda p_i}\, ,
        \end{equation}
        when the partial pressure $p_{i}$ satisfies:
        \begin{equation}
        \label{equ:limit}
            p_i < \frac{1}{I_{\sigma}\lambda} \, .
        \end{equation}
        As $p_i$ approaches this limit, the number of generated He ions in a PMT event is such that the individual after-pulses have a similar integrated charge to the primary pulse. Each successive ionisation stage is sufficient to produce a significant amount of charge which can result in a continuous discharge-like breakdown of the PMT. It is likely that significant after-pulsing will be evident long before this limit as a possible sign of degradation.   
    
    \subsection{After-pulse Times}
    \label{subsec:apulsetime}
        In general, the drift time of the ionised gas atoms/molecules will depend on where the ionisation event takes place and the geometry of the electric field within the PMT. The case of a hemispherical PMT, which is a good approximation for the photomultiplier tubes used in this study, was analysed in~\cite{Incandela:1987dh}. A particular feature of this geometry is that the drift time is approximately independent of the position of the initial ionisation, and is given by:
        \begin{equation}
            \label{equ:timeofflight}
            \tau \approx \frac{\pi}{2}\sqrt{\frac{md^2}{2ZeV_0}} \; ,
        \end{equation}
        where $m$ is the mass of the charged ion species with charge $Ze$ and $V_{0}$ is the potential difference across the hemispherical volume with a distance $d$ between the photocathode and the dynode chain. For example, the PMTs that are the subject of this study described in Section 3 operated at $1400$~${\rm V}$ with the voltage divider used in the SuperNEMO experiment, $V_{0}$ = $600$~${\rm V}$ as the voltage drop between the photocathode and the first dynode and $d = (10\pm1)$~${\rm cm}$ to give an expected $\tau_{\rm He^{+}} = (0.9\pm0.1)$~${\rm
        \mu s}$. While this provides an initial target for an after-pulse search in our data, it is expected that the after-pulse time distribution will be shifted and broadened with respect to this estimate for a number of reasons including inexact knowledge of the field-map inside the tube and the possibility of ionisation events taking place in regions of low-field or deeper within the dynode stages. Finally, it should be noted that equation~\ref{equ:timeofflight} allows, in principle, the identification of atomic and molecular species of a given charge-to-mass ratio $(Z/m)$ based on the after-pulse time, in direct analogy with mass-spectroscopic techniques.
        
    \subsection{Helium Permeation Model}
        \label{subsec:he_permeation}
        The atomic structure of glass gives rise to pathways through which atoms can diffuse; this is especially pronounced for He which has a small atomic radius and is chemically inert. However, the presence of non-glass forming oxides within the bulk glass material can significantly reduce the permeability of glass to He, thus extending a PMT's lifetime.
        \par 
        Typical permeation rates for He through different glasses can be found in \cite{Altemose}. The SuperNEMO R5912-MOD PMT glass has a borosilicate classification with a percentage of glass forming oxides below 90\%, making it significantly less permeable than, for example, certain fused-silica glass types. The relationship between the percentage of glass forming oxides, the activation energy and the permeation constant is explored in Incandela et al~\cite{Incandela:1987dh}. In this study, the helium diffusion constant, $D$, and the solubility of helium in glass, $S$, which together determine the permeability of the glass, $K$, are not calculated \textit{a priori}. Instead, an empirical model is employed to fit the data such that the parameters can be extracted.
        \par
        For a PMT volume, $V$, with a glass of thickness $l$, that is exposed to an external partial pressure of He, $p_e$, the increased internal partial pressure of He, $p_i$, within the volume that has diffused through the glass is given by \cite{Fick}: 
        \begin{equation}
        \label{equ:pi}
            \begin{split}
                p_i(t) =& \frac{DSA}{l}\frac{RT}{V}p_e(t - L) - \frac{12L}{\pi^2}\frac{DSA}{l}\frac{RT}{V}p_e\sum^{\infty}_{n=1}\frac{(-1)^n}{n^2}e^{\frac{-n^2\pi^2t}{6L}} \\
                =& \Gamma_{i}p_e(t - L) - \frac{12L}{\pi^2}\Gamma_{i}p_e\sum^{\infty}_{n=1}\frac{(-1)^n}{n^2}e^{\frac{-n^2\pi^2t}{6L}} 
                \\
                =& \mathcal{A}(t-L) + \mathcal{B}(t)
            \end{split}
        \end{equation}
        where $t$ is time from onset of exposure, $R$ is the universal gas constant, $T$ is the ambient temperature and $A$ is the surface area of the PMT bulb. This equation combines a linear term for steady state diffusion, $\mathcal{A}(t-L)$, and a higher order term which describes the initial behaviour of the gas interacting with the glass, $\mathcal{B}(t)$. The lag time is represented by the parameter $L=\frac{l^2}{6D}$, and is typically hours or days for practical PMTs~\cite{Incandela:1987dh, Altemose}. The parameter $\Gamma_i$ describes the general diffusion rate in~$\text{s}^{-1}$.  
        
    \subsection{Measuring the Internal Partial Pressure of Helium}
    \label{subsec:He_pi}
        Directly measuring the internal partial pressure of He within a PMT bulb is difficult and impractical as it would likely require breaking the PMT. Therefore, two proxy methods of estimating $p_i(t)$ using the observable after-pulses have been developed. The model in equation \ref{equ:pi} is then used to fit this proxy $p_i$ to determine values for the parameters $\Gamma_i$ and $L$ so that PMT lifetime estimates can be made.

        \subsubsection{Proxy A: After-pulses}
        \label{subsubsec:proxyAapulses}
            $\text{He}^+$ ions that form in the PMT volume produce observable after-pulses whose time delay after the initial pulse is governed by equation \ref{equ:timeofflight}. Assuming that each He ion produces one after-pulse, equation \ref{equ:N_int_x} can be used to estimate $p_i$ using the measured number of after-pulses, $A_n$, as follows: 
            \begin{equation}
            \label{equ:apulse_pi_recon}
                p_i = \frac{N_{\text{He}^+}}{I_{\sigma}n_{\text{p.e}}} = \frac{eG_{\rm PMT}}{I_{\sigma}}\frac{A_n}{Q_0}=\frac{eG_{\rm PMT}}{I_{\sigma}}R_{n},
            \end{equation} 
            where we use the fact that the measured primary pulse charge, $Q_0$, in a PMT is a product of the initial charge of the photoelectrons, $n_{\text{p.e}}$, the charge of the electron and the gain, $G_{\rm PMT}$. The after-pulse number, $A_n$, will also vary with $Q_0$, therefore, the ratio, $R_n = \langle A_n\rangle/\langle Q_0\rangle$ is the measurable parameter. 

        \subsubsection{Proxy B: He After-pulse Region Charge}
        \label{subsubsec:proxyBapcharge}
            Measuring the number of after-pulses is challenging as it depends on the size of the smallest after-pulse. Another proxy method to measure $p_i$ is using the integrated charge in the region of the PMT waveform, after the primary PMT pulse, where after-pulses caused by the $\text{He}^+$ species are expected to occur. This region is governed by equation \ref{equ:timeofflight}, and is referred to as the He after-pulse region in this paper and is defined later in section \ref{sec:afterpulse-dists}. The He after-pulse region charge, $Q_1$, is directly proportional to the primary pulse charge, $Q_0$. The ratio, $R_c = \langle Q_1/Q_0\rangle$, is another measurable parameter to determine $p_i$ as follows:
            \begin{equation}
            \label{equ:lambda_pi}
                p_i = \frac{N_{\text{He}^+}}{I_{\sigma}n_{p.e}} = \frac{1}{I_{\sigma}\lambda} \frac{n_{\text{s.e}}}{n_{p.e}} = \frac{1}{I_{\sigma}\lambda} \frac{Q_1}{Q_0} =\frac{1}{I_{\sigma}\lambda}R_c,
            \end{equation}
            where the substitution $N_{\text{He}^+} = n_{\text{s.e}}/\lambda$ has been made. Note that the ratio of the numbers of electrons in the primary and secondary pulses is equivalent to the ratio of their charges, with the PMT gain canceling in the ratio.

\section{Experimental Setup and Analysis Techniques}
\label{sec:experiment}
    \subsection{SuperNEMO Photomultipler Tubes}
        \label{sec:SuperNEMO_PMTs}
      
        The SuperNEMO calorimeters are mainly instrumented with Hamamatsu 8" photomultiplier tubes of type R5912-MOD. These have a reduced number of dynode stages (8) due to the high light yield of typical calorimeter hits in SuperNEMO, and have a gain of approximately $10^{5}$ at $1400$~${\rm V}$. The voltage dividers attached to the PMT bases have been optimised in previous studies; the total impedance of the divider is $8.06$~${\rm M\Omega}$, and the voltage difference between the photocathode and the first dynode is $42\%$ of the applied high-voltage. At $1400$~${\rm V}$, this corresponds to approximately $600$~${\rm V}$. Further details can be found in~\cite{wq_thesis}.
        
    \subsection{PMT Test-Stand}
    \label{subsec:PMT Test-Stand}
\begin{figure}[h!]
            \centering
            \includegraphics[scale=0.3]{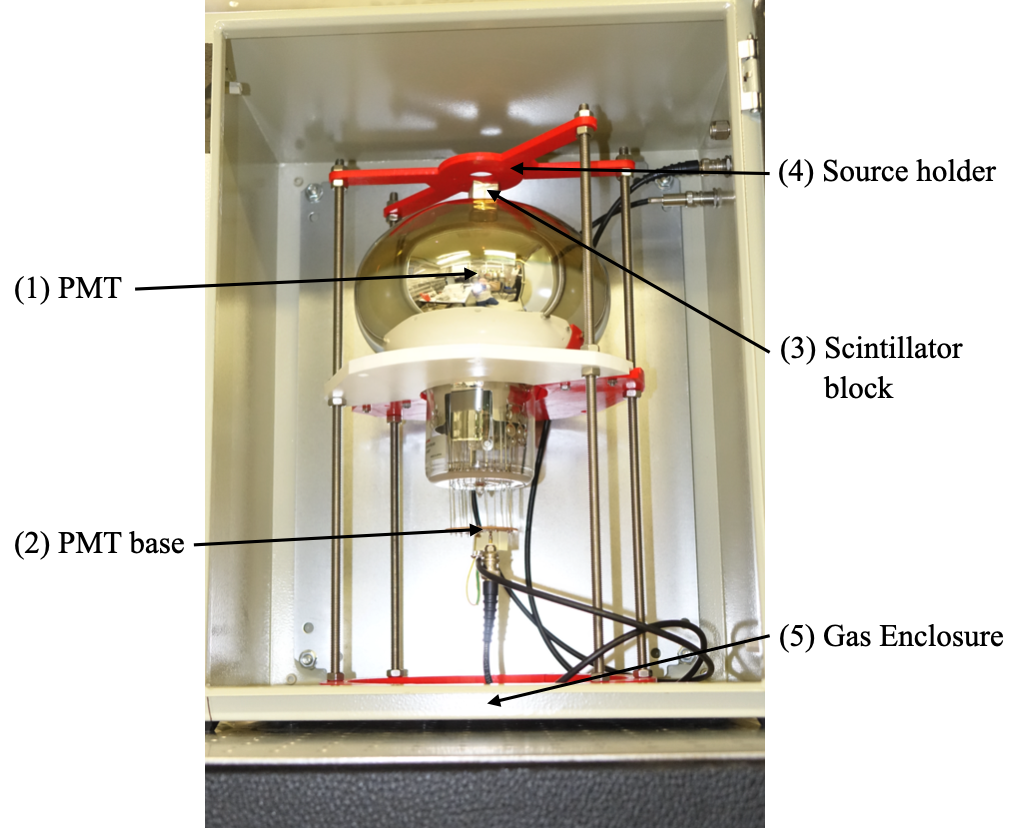}
            \caption{The R5912-MOD Hamamatsu PMT (1) and electronics base (2) in a holder with a small scintillator block (3), wrapped in Mylar, optically coupled to the PMT glass. The source holder (4) holds a small \nucl{207}{}{Bi} radioactive source close to the scintillator block. The PMT rig is placed in a container (5) that is made light tight.}
            \label{fig:PMT}
        \end{figure}

        Two R5912-MOD Hamamatsu PMTs were placed in light-tight containers, one exposed to He and the other monitored as a control in nitrogen. The exposed container was made gas tight with an input-output feed for gas injection. A slow gas flow rate was maintained to ensure a constant concentration of He at the desired level. The PMTs were both optically coupled to small plastic scintillator cubes measuring 6.4 cm in width and placed in close proximity to separate \nucl{207}{}{Bi} sources with activities of $\sim$30 kBq. A photograph of one of the PMTs is shown in figure~\ref{fig:PMT}.
        \par
        Data was taken over one year beginning with atmospheric He conditions. A 1\% He, 99\% nitrogen mixture was then flowed for 98 days after which no significant resolution degradation or after-pulse affect was observed, as shown in section \ref{sec:energy_res}. The mixture was then changed to 10\% He, 90\% nitrogen and flowed for the remainder of the experiment.
        
        \subsubsection{Data Acquisition}
            A CAEN digitiser (model DT5751 \cite{CAEN}) was used to acquire the PMT signals. The dynamic range was set to 1~V with a granularity of 1~mV as to maximise the sensitivity to small PMT pulses, typical of after-pulses. The number of ADC samples taken per waveform was set to 7000 at a frequency of 1~GHz, corresponding to a total waveform duration of $7$~${\rm \mu s}$.
            \par
            For the main after-pulse study the PMT voltage was set to $1400$~V, representing the running conditions used in SuperNEMO. However, unlike the Demonstrator electronics, at this voltage the large-light pulses saturate the test-stand ADC, therefore a Time-Over-Threshold (TOT) analysis is used to reconstruct the charge of the primary pulses, $Q_0$. Hour-long runs were taken at 1000~V to reconstruct the \nucl{207}{}{Bi} spectrum without saturation, which is important for analysing the impact of He exposure on energy resolution. The PMT gain at 1000~V in the energy region of $1$~${\rm MeV}$ is estimated using the fitted resolution. This is then scaled up to get the effective PMT gain (which may also incorporate a change in the photoelectron collection efficiency) at 1400~V by the ratio of the fitted centroid of $Q_0$ for the $1$~${\rm MeV}$ peak at 1400~V and 1000~V. This varies across the course of the experiment by $\sim$5-10\% and is shown in figure \ref{fig:1meV_data}.
        
    \subsection{After-pulse Finding}
        \label{subsec:apulse_finding}
        A Matched Filtering (MF)\cite{HUANG2005811} convolution analysis was applied to determine the number of PMT after-pulses within the noise of the PMT waveform. The output of the MF convolution enables a more sensitive after-pulse search than a simple threshold-based pulse identification as it maximises the signal to noise ratio. The MF convolution produces two continuous outputs: the \textit{shape} and the \textit{amplitude}. The \textit{shape} represents the cosine similarity between a known signal template and the corresponding segment of the waveform, while the \textit{amplitude} represents the scaled cosine similarity, proportional to the size of the waveform segment. Peaks within these outputs indicate potential signal candidates within the waveform, characterized by specific \textit{shape} and \textit{amplitude} indices.
        \begin{figure}[h!]
            \centering
            \includegraphics[scale=0.99]{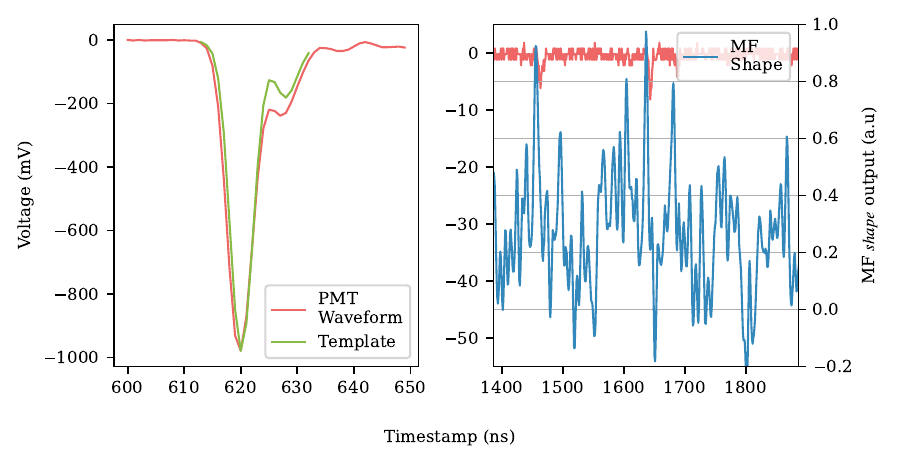}
            \caption{Left: example PMT pulse template (green) scaled and overlaid onto a PMT pulse signal (red). Right: zoomed in view of the corresponding PMT post primary pulse (after-pulse) region (red), alongside the MF convolution \textit{shape} output. It is clear that peaks in the \textit{shape} output correspond to pulse-like samples within the post primary pulse region.}
            \label{fig:convolution}
        \end{figure}
        \par
\begin{figure}[h!]
            \centering
            \includegraphics{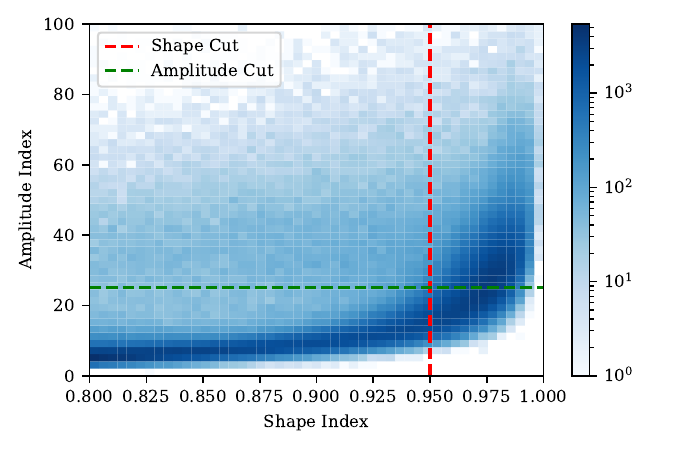}
            \caption{Example MF convolution output distribution from a $10^{5}$ waveform sample. The chosen cuts to define an after-pulse are displayed, \textit{shape} index of $>$0.95 (red-dashed) and an \textit{amplitude} index of $>$25 (green-dashed).}
            \label{fig:cuts}
        \end{figure}

 
        An example distribution of these MF peak indices for a data set can be seen in figure \ref{fig:cuts} from which cuts of \textit{shape} $>0.95$ and \textit{amplitude} $>25$ were applied to define an after-pulse. An \textit{amplitude} index of 25 corresponds to a voltage amplitude of 12~mV at 1400~V which in turn corresponds to approximately 5.4 photoelectrons. The choice of these cuts was estimated by injecting template pulses, whose amplitude followed a uniform distribution between 1-100 mV, onto baseline data and performing a MF convolution. The cuts were optimised by minimising the false positive rate of finding the injected pulses whilst maximising the true positive rate. The overall efficiency was determined by the survival fraction of the injected waveforms, which was found to be 73\%. However, this value has a significant systematic uncertainty since the true distribution of the amplitude of the after-pulses is not known without first applying a set of cuts; the results of the MF analysis remain sensitive to the choice of matching cuts, as discussed in section~\ref{subsec:he_modelling}.
        \par
\begin{figure}[h!]
            \centering
            \includegraphics{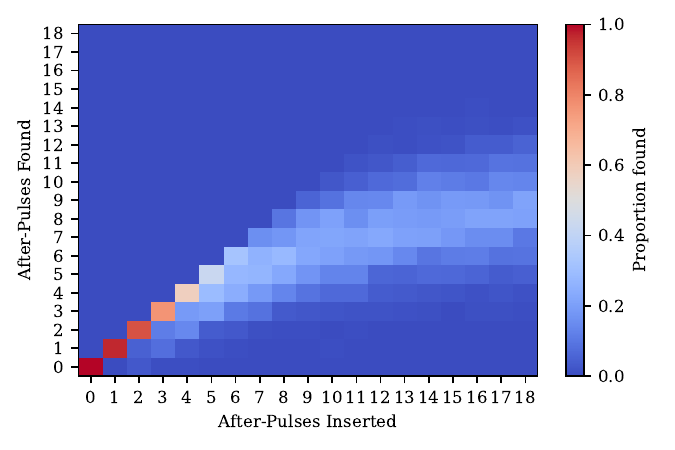}
            \caption{The after-pulse pile-up MF inefficiency matrix showing the relationship between measured and inserted after-pulse numbers. PMT pulse templates were inserted into 500 sampled baseline waveforms, for each instance on the x-axis, and a MF convolution was employed to find them within the baseline noise, as discussed in the text. The number of pulses found in a baseline, for a given injected number, is populated in the z-axis normalised by the sample size of 500.}
            \label{fig:matrix}
        \end{figure}

        Since MF uses shape information, a potential source of inefficiency arises when multiple after-pulses overlap with offset start-times, giving rise to single pulses with different shapes. 
        To quantify this inefficiency, a simulation was performed by injecting multiple after-pulse template pulses into realistic baselines. These were injected following the measured time distribution, such as those in figure \ref{fig:apulsetimes}, and with a MF \textit{amplitude} cut > 25. Comparing the number of injected pulses to the number found using the MF convolution yields the inefficiency which is represented by the off-diagonal elements in the matrix shown in figure~\ref{fig:matrix}. The mean of the measured after-pulse number, $\langle A_n\rangle$, is scaled by the cut efficiency and corrected using this matrix to obtain a better estimate of the true after-pulse multiplicity, $\langle A_n\rangle^{(c)}$~\cite{wq_thesis}.

\section{Results and Analysis}
\label{sec:ap_rates}
    \subsection{Energy Resolution}
    \label{sec:energy_res}
\begin{figure}[h!]
            \centering
            \includegraphics[scale=1.]{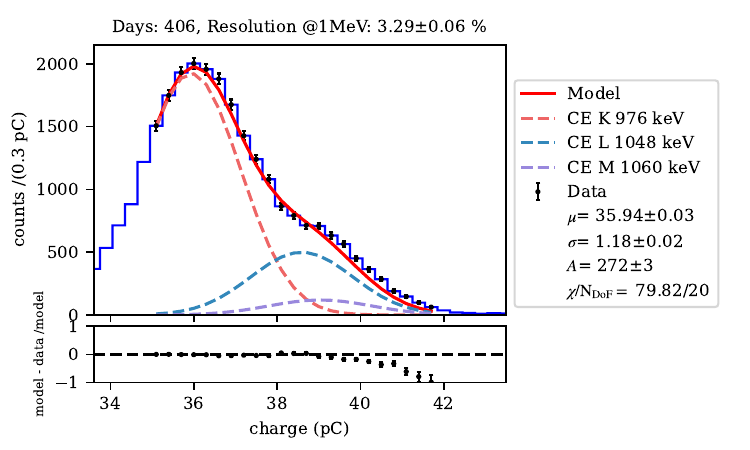}
            \caption{Top: PMT \nucl{207}{}{Bi} 1 MeV conversion electron charge distribution fit with a functional form described in the text. The fitted parameters are displayed where $A$ is an arbitrary overall normalisation factor. Bottom: data/model residual comparison.}
            \label{fig:bi_fit}
        \end{figure}

\begin{figure}[h!]
            \centering
            \includegraphics[scale=1.]{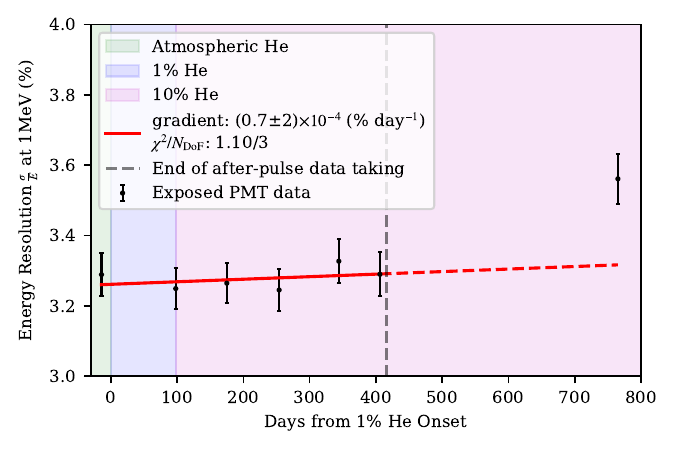}
            \caption{Calculated energy resolution, in $\sigma$, at 1 MeV of R5912-MOD Hamamatsu PMT exposed to varying external He exposures as a function of time. The red fitted straight line shows a gradient consistent with zero over 300 days of 10\% He exposure. The dashed red line is an extrapolation of the fit to larger exposures.}
            \label{fig:res}
        \end{figure}

        The energy resolution is the chosen metric to determine the evolving PMT performance as a function of He exposure. The resolution, defined as the ratio of integrated primary peak charge, $Q_0$, width (in $\sigma$) to peak position at 1~MeV, was calculated for both PMTs using the $K_{976}$, $L_{1048}$ and $M_{1060}$ conversion electron peaks of the \nucl{207}{}{Bi} isotope \cite{KONDEV2011707}. Using the lower PMT voltage of 1000 V, the collected pulse charge spectra were fitted using three summed Gaussian distributions, one for each conversion electron, as shown in figure \ref{fig:bi_fit}. This is a simplified continuum where the underlying Compton background is neglected with no detriment to the determination of the energy resolution \cite{wq_thesis}. This procedure was then repeated for other sample 1000~V datasets taken across the timescale of the experiment. The measured resolution is displayed in figure \ref{fig:res} along with a linear functional fit. For over 300 days exposure to 10\% He, as well as 100 days at 1\%, the exposed PMT energy resolution did not degrade significantly while the after-pulse rate increased, indicated by the fitting output for the gradient,$(0.7\pm2)\times10^4~\%~\text{day}^{-1}$, being consistent with zero.
        \par 
        The PMT was then exposed to 10\% for approximately another 300 days (approximately 650 days after the onset of 10\% He) to see how long it could last before any noticeable breakdown occurred. The last data point in figure \ref{fig:res} shows evidence of mild degradation in energy resolution when compared to the extrapolation of the linear fit to the earlier energy resolution data points. Overall, a negligible impact of He permeation on energy resolution can be seen, even in the presence of strong after-pulsing behaviour.
        \par 
    \subsection{After-pulse Distributions}
    \label{sec:afterpulse-dists}
\begin{figure}[h!]
            \centering
            \includegraphics{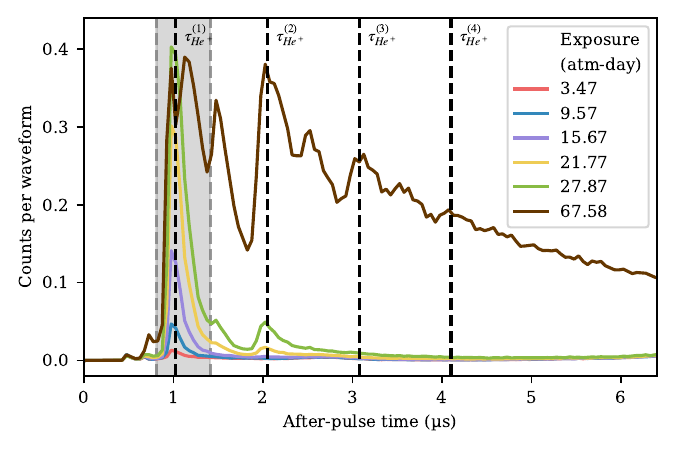}
            \caption{R5912-MOD PMT measured after-pulse times ($\tau$) for different exposures of He, normalised to the number of collected waveforms, using a MF pulse finding algorithm. The predicted value of the after-pulse time for the He$^+$ species is shown at $\tau^{(1)}_{\rm He^+}=0.9~\mu$s, as well as the recursive after-pulse times, $\tau^{(i)}_{He^+}$, at $0.9~\mu$s intervals. The He after-pulse region, (810-1410)~ns, is shown in grey.}
            \label{fig:apulsetimes}
        \end{figure}

\begin{figure}[h!]
            \centering
            \includegraphics{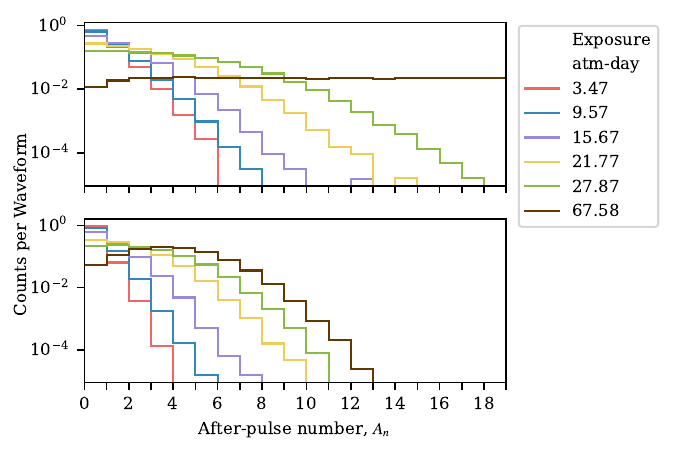}
            \caption{R5912-MOD PMT measured after-pulse number for different exposures to He, normalised to the number of collected waveforms. Top: total after-pulse number in the entire waveform post primary PMT pulse. Bottom: after-pulse number in the He after-pulse region, (810-1410)~ns, shown in figure \ref{fig:apulsetimes}.}
            \label{fig:ap_num_dist}
        \end{figure}

        The after-pulse datasets were acquired daily at a higher voltage of 1400~V. This maximised after-pulse signals but the high-energy primary pulses sometimes exceed the dynamic range of the DAQ. The after-pulse times were determined using a MF and compared to the primary pulse time\footnote{Due to saturation the estimated pulse start time has a systematic uncertainty of $\pm20$ ns.}. These after-pulse time distributions for different He exposures can be seen in figure \ref{fig:apulsetimes}. Furthermore, the distributions of the number of found PMT after-pulses for the same exposures are shown in figure \ref{fig:ap_num_dist}. The average of each of these distributions is calculated then corrected by the MF inefficiencies for use in the data proxy introduced in section \ref{subsubsec:proxyAapulses}.
        \par 
         The features within figure \ref{fig:apulsetimes} appear to grow with He exposure which is indicative of induced after-pulsing. Specifically, the feature at 1 $\mu$s appears the strongest which is assumed to be the He$^+$ ion species, the most likely candidate, and close to the predicted value of ($0.9\pm0.1$)~$\mu$s. At extreme exposures the effect of recursive after-pulsing, discussed in section \ref{subsec:int_p_limit}, can clearly be seen at 1 $\mu$s intervals, denoted by $\tau^{(i)}_{He^+}$ in figure \ref{fig:apulsetimes}. From these distributions a time window in the PMT waveforms is defined to apply the proposed $p_i$ reconstruction methods, as described in section~\ref{subsec:He_pi}.

    \subsection{Determining $\lambda$}
    \label{sec:lamda}
\begin{figure}[h!]
            \centering
            \includegraphics{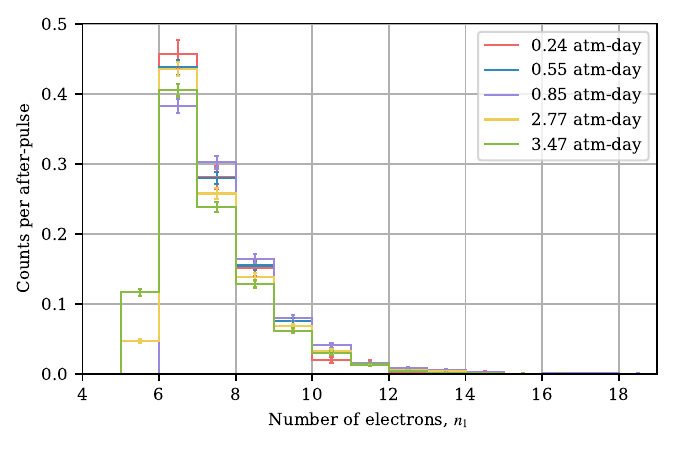}
            \caption{Reconstructed number of electrons released from the photocathode by He ions in a single after-pulse event for increasing He exposures. These were obtained using the measured \textit{amplitude} index of the found after-pulse from the MF and the template unit charge.}
            \label{fig:ap_pe}
        \end{figure}

\begin{figure}[h!]
            \centering
            \includegraphics{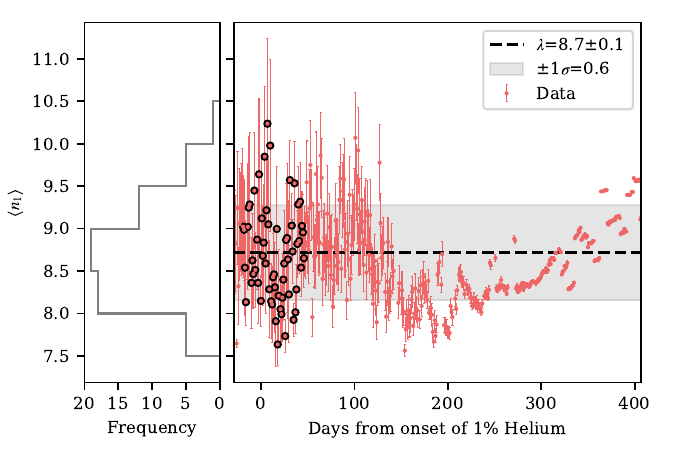}
            \caption{Right: Average number of electrons of found after-pulses with example distributions shown in figure \ref{fig:ap_pe}. The mean, error on the mean and the standard deviation are displayed. Left: y-axis projection of the selected data points highlighted with a black edge.}
            \label{fig:ap_pe_vs_time}
        \end{figure}

        The average number of electrons released per He ion collision is referred to as $\lambda$, and is required to reconstruct the internal partial pressure of He inside the PMT bulb. Using the after-pulses found by the MF method, the charge of each after-pulse is calculated using its \textit{amplitude} index multiplied by the unit charge of the template after-pulse, which is then divided by the PMT gain to reconstruct the number of electrons that are released from the photocathode, $n_1$. This is shown in figure \ref{fig:ap_pe} for increasing He exposures. The sharp cut off at lower values of $n_1$ is due to the applied MF \textit{amplitude} cut of $>$25, which corresponds to $n_1\approx5.4$. Taking the average of this distribution provides an estimate for $\lambda$. It was decided to only use datasets with minimal He exposures to minimise the effect of the MF method pile-up inefficiency that increases at higher exposures. This is shown in figure \ref{fig:ap_pe_vs_time}. The calculated value of $\lambda$ is $8.7\pm0.1_{\text{stat.}}$, using the error on the mean as the uncertainty. This value is sensitive to the chosen after-pulse cuts defined earlier so the standard deviation of $\langle n_1\rangle$ was included as a systematic uncertainty though it is likely that $\lambda$ is biased to a higher value.
            
    \subsection{Modelling Results}
    \label{subsec:he_modelling}
\begin{figure}[h!]
            \centering
            \includegraphics{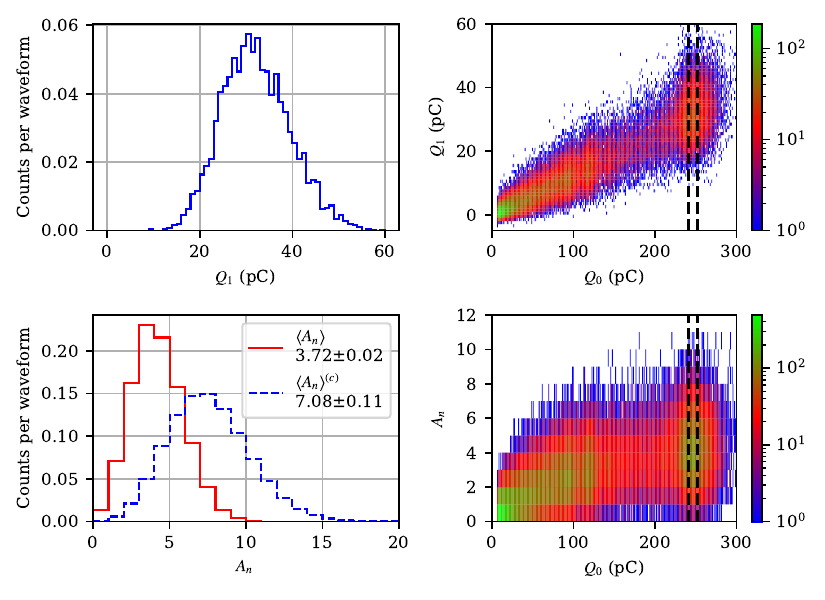}
            \caption{Measured He after-pulse region (890-1410 ns) distributions at 32.27 atm-day exposure of He. Top-left: The He after-pulse region charge, $Q_1$, for PMT pulse charges, $Q_0$, in the \nucl{207}{}{Bi} 1~MeV peak shown in the top-right. Top-right: $Q_1$ as a function of $Q_0$, the 1~MeV \nucl{207}{}{Bi} peak is displayed as the dashed vertical lines at approximately 240~pC. Bottom-left: The number of found after-pulses in red for PMT pulse charges, $Q_0$, at the \nucl{207}{}{Bi} 1~MeV peak. Also displayed is the Poisson distribution, in blue, which was generated by correcting the average of the data in red, via the method described in the text. Both averages, $\langle A_n\rangle$ and $\langle A_n\rangle^{(c)}$, are displayed with statistical errors. Bottom-right: $A_n$ as a function of $Q_0$, the 1~MeV \nucl{207}{}{Bi} peak is displayed as the dashed vertical lines at approximately 240~pC.}
            \label{fig:1meV_dist}
        \end{figure}

\begin{figure}[h!]
            \centering
            \includegraphics{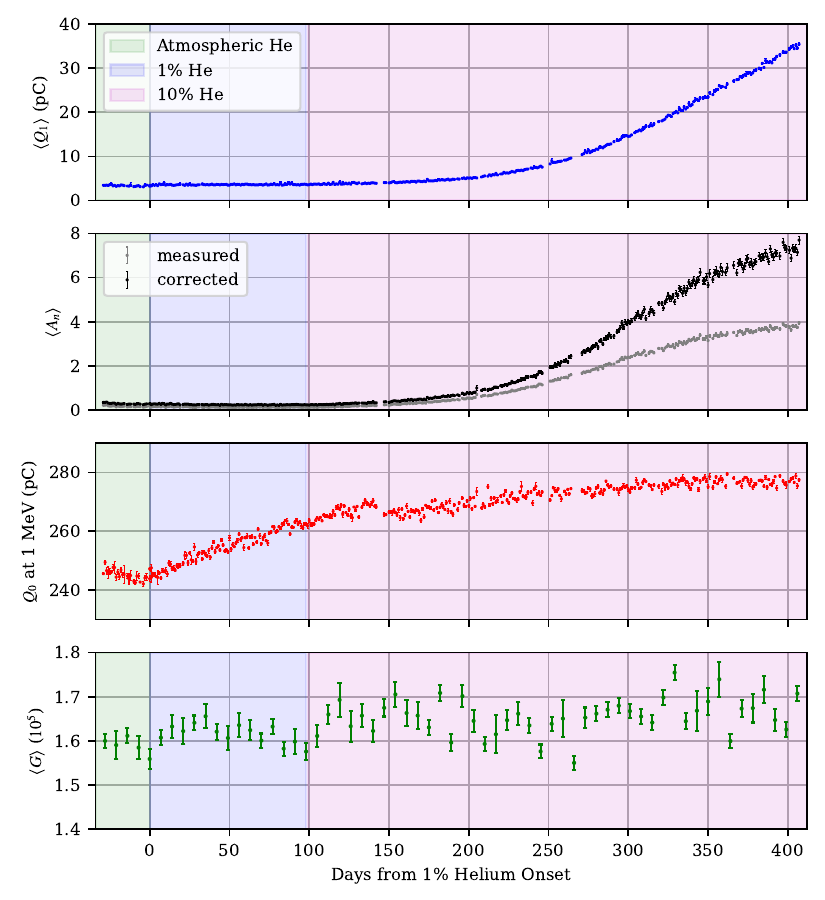}
            \caption{1st Panel: The average He after-pulse region (810-1490 ns) charge, $\langle Q_1\rangle$ as a function of time. 2nd Panel: The measured He after-pulse region average after-pulse number, $\langle A_n\rangle$, and the corrected average, $\langle A_n\rangle^{(c)}$ as a function of time. 3rd Panel: The PMT pulse charge, $Q_0$, of the \nucl{207}{}{Bi} 1 MeV peak as a function of time. 4th Panel: The Reconstructed PMT gain at a PMT voltage of 1400 V as a function of time. The time periods of different external He partial pressures are also displayed.}
            \label{fig:1meV_data}
        \end{figure}

        Figure \ref{fig:1meV_dist} shows how the after-pulse number and the He after-pulse region charge, the proxies introduced in sections \ref{subsubsec:proxyAapulses} and \ref{subsubsec:proxyBapcharge}, respectively, vary as a function of energy at an exposure of 32.27 atm-day (day 410 from onset of 1\% He). The projection of these distributions for PMT pulses around the 1~MeV \nucl{207}{}{Bi} peak ($\pm$5~pC) is also shown in figure \ref{fig:1meV_dist}. The average of these distributions, the corrected average after-pulse number, $\langle A_n\rangle$, and the average He after-pulse region charge, $\langle Q_1\rangle$, are shown as a function of time in figure \ref{fig:1meV_data}. Also displayed are the variation of the centroid of the charge of the 1~MeV peak and the weekly average of the PMT gain at 1400~V. 
        \par
\begin{figure}[h!]
            \centering
            \includegraphics{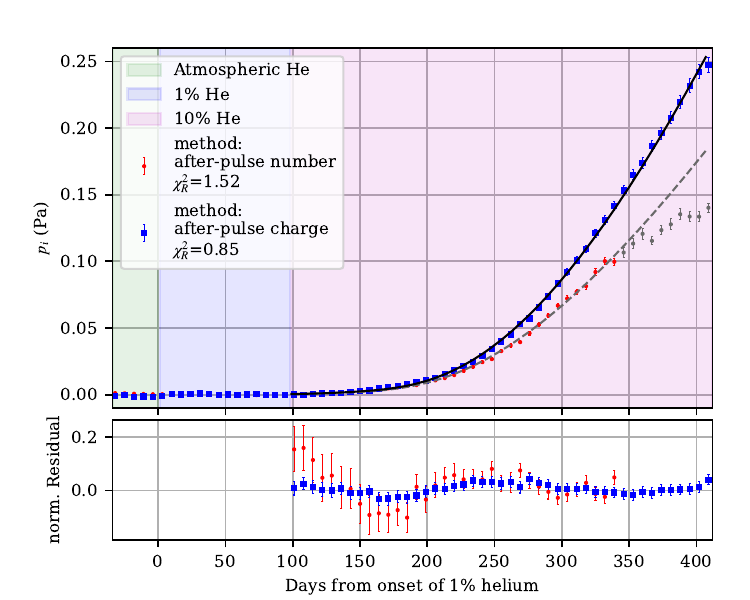}
            \caption{The reconstructed internal partial pressure of He, $p_i$, from the two methods described in the text. Top: The best fit models and data are displayed with the corresponding $\chi_R^2$ whose parameters are displayed in table \ref{tab:model_pars_transposed}. The time periods of different external He partial pressures are also displayed. Bottom: the residual of the model to the data scaled by the model.}
            \label{fig:recon_pi}
        \end{figure}

        The reconstructed $p_i$ for both the corrected average after-pulse number and He after-pulse region charge methods are shown in figure \ref{fig:recon_pi} where only PMT pulses within the 1~MeV \nucl{207}{}{Bi} peak ($\pm$5~pC) of the 1400 V spectrum were considered\footnote{Although data was taken daily, the weekly average has been computed and displayed for better visualisation.}. In both cases a series of models were fitted and the ones with the best $\chi^2_R$ are displayed. These models are small variations of equation \ref{equ:pi} thus share the same set of parameters whose spread can be used as a systematic uncertainty. The best model is found to be an extended model that incorporated contributions, expressed by equation \ref{equ:pi}, from both the 98 day period of 1\% He and the 300 days at 10\% He exposure \cite{wq_thesis}. The parameter output, the general diffusion rate $\Gamma_i$, the lag time, $L$, and the effective constant offset, $p_1$, for each method are shown in table \ref{tab:model_pars_transposed} with their associated statistical uncertainty that is derived from the fit error. 
        \begin{table}[htbp]
            \centering
            \caption{The fitted parameters; the general diffusion rate, $\Gamma_i$, the lag time, $L$ and the effective constant offset, $p_1$; for the He permeation model, defined in equation \ref{equ:pi}. Parameters fitted to the reconstructed internal partial pressure of He within the R5912-MOD PMT, $p_i$, for both the proxy methods discussed in the text; proxy A, the average after-pulse number introduced in section \ref{subsubsec:proxyAapulses}, and proxy B, the He after-pulse region charge introduced in section \ref{subsubsec:proxyBapcharge}. The model and the data are shown in figure \ref{fig:recon_pi}. The errors reported here are purely statistical taken from the result of the fit.}
            \begin{tabular}{c|c|c}
                & A: After-pulse & B: Charge\\
                \hline
                \hline
                $\Gamma_i$ $(10^{-12}\text{s}^{-1})$ & $1.68\pm0.01$ & $2.67\pm0.01$  \\
                $L$ (days)                           & $236\pm1$     & $263\pm1$      \\
                $p_1$ (mPa)                          & $5.0\pm0.1$   & $32.0\pm0.1$ \\
                \hline
                $\chi_R^2$                           & $1.52$        & $0.74$         \\
            \end{tabular}
            \label{tab:model_pars_transposed}
        \end{table}
        \par
        An extra parameter, $p_1$, is included as a constant offset in both proxy methods and has been subtracted from the data presented in figure~\ref{fig:recon_pi}. $p_1$ varies significantly between the two methods; despite a simple interpretation as the primordial He pressure inside the PMT, in practice it accounts for all contributions to the baseline charge and after-pulse counts in the two methods. Our results and modelling only seek to capture the {\em additional} He partial pressure arising from diffusion through the PMT glass. 
        \par
        With the onset of 10\% He exposure, the reconstructed $p_i$ increases significantly as the after-pulse number rises, in both scenarios. The permeation model is a good fit though there is clear evidence of higher order effects not addressed by the chosen \textit{extended} model. Taking into account the systematic uncertainties on $I_{\sigma}$ (3\%) and $\lambda$ (6\%), $\Gamma_i=~(2.70\pm~0.01_{\text{stat.}}\pm~0.18_{\text{sys.}})\;10^{-12}~\text{s}^{-1}$.
        \par 
        The attempts to correct the number of measured after-pulses is clearly not sufficient as highlighted by the plateauing effect at larger pulse charges in figure \ref{fig:1meV_dist} and in larger exposures in figure \ref{fig:recon_pi}. In an attempt to reduce this effect on the fitting output, larger exposures have been neglected from the fit as shown by the greyed-out data points in figure \ref{fig:recon_pi}. The integrated charge method appears to have less deviation from the model and is intrinsically a simpler measurement, therefore, it is the method and model chosen to use to make predictions of PMT lifetime later on in this paper. The after-pulse number method has only been used as a cross-check and to obtain after-pulse time information.
        \par
        The fitted value for $\Gamma_i$ can be used to calculate a value for the permeation constant of He through glass, $K$. The PMT is assumed to be spherical with a radius of (10$\pm$1) cm and an average glass thickness of (2$\pm$1)~mm at room temperature and pressure. This yields a value of $(0.754\pm~0.003_{\text{stat.}}\pm~0.05_{\text{sys.}}\pm~0.380_{\text{geom.}})~\times~10^{-20}~\text{mol}~\text{mm}~\text{cm}^{-2}~\text{Pa}^{-1}~\text{s}^{-1}$, where the larger systematic uncertainty due to the geometry assumptions of the PMT has been separated out. The value of $K$ can be cross-checked with the empirically deduced formula taken from \cite{Altemose}. The percentage of glass forming oxides, $M$, for the R5912-MOD PMT is used \cite{Oxides}, from which at 293~K the value of $K$ is calculated to be $(2.1\pm0.7_{\text{sys.}})\times10^{-20}~\text{mol}~\text{mm}~\text{cm}^{-2}~\text{Pa}^{-1}~\text{s}^{-1}$. These values do not agree at 1$\sigma$ however; both have large sources of systematic uncertainty.
        \par
        Furthermore, the lag time parameter, $L$, can also be compared to the value predicted using the method discussed in \cite{Incandela:1987dh}. The value of $L$ extracted from modelling and the predicted range are $263\pm1$ and 64-240 days, respectively, where the uncertainty is taken from the fit. The degree of agreement is hard to distinguish within the large uncertainties.
        
    \subsection{Lifetime Estimates}
        It is clear from the constructed pressure limit in equation \ref{equ:limit}, that the value of $\lambda$, the average number of electrons released per He ion collision, is not required to assess the lifetime of a PMT; one might simply measure and model the ratio of the secondary and primary pulse charges, $R_c$, and extrapolate to when $R_c=1$. However, using equation \ref{equ:limit}, the pressure limit calculated with $\lambda=8.7$, is $P_{\text{lim}}=3.13$ Pa. There is little in the relevant literature that discusses the internal partial pressure break-down point. One source from ET Enterprises \cite{ETEnterprises} quotes a level of 0.001~Torr (0.133 Pa), for considerable after-pulse rates rending the PMT inoperable. Using the fitted parameters, a range of times, $T_{\text{lim}}$, are presented to reach 3.13 Pa in different He exposure conditions in table \ref{tab:predictions}.
        \begin{table}[h!]
            \centering
            \caption{The estimated Hamamatsu PMT lifetimes for given exposures of He concentrations relative to atmospheric pressure for a PMT breakdown pressure of 3.13~Pa.}
            \begin{tabular}{cc}
                $p_e$ (\% atm) & $T_{\text{lim}}$ (yrs) \\
                \hline
                10 & 4.4 \\
                1 & 37 \\
                0.1 & 370 \\
                \hline
                
            \end{tabular}
            \label{tab:predictions}
        \end{table}
        
\section{Conclusions}
    An investigation into the He poisoning of Hamamatsu 8" R5912-MOD PMTs, used in the Demonstrator module of the SuperNEMO neutrinoless double-beta decay experiment, has been undertaken. Two PMTs, one exposed to 1\% He for 100 days and 10\% He for two years, and one kept in atmospheric conditions, were monitored for performance changes and after-pulse rates. The energy resolution of optical modules formed with scintillator blocks attached to the the PMTs was measured using a \nucl{207}{}{Bi} radioactive source and was found to be stable over 300 days in 10\%~He. There is a clear increase in the rate of after-pulses in the exposed PMT, with a time distribution in good agreement with the drift time expected for He ions, and with evidence of secondary and tertiary after-pulse populations. 
    \par
    The most robust method of reconstructing the after-pulse behaviour is to measure the charge in a time-interval following the primary pulse where He after-pulsing is present. A cross-check using a MF based pulse shape analysis to count the after-pulses is in approximate agreement with the charge integration method, but subject to larger systematic uncertainties. In both cases the data show good agreement with He permeation theory, enabling the lag-time and diffusion constant to be extracted.
    \par
    A partial pressure of He at which the PMTs would become inoperable is estimated to be 3.13~Pa. At this pressure the after-pulse charge in generation $i$ would become similar to the after-pulse charge in subsequent generation $i+1$, leading to a quasi-continuous breakdown. For the PMT exposed to 10\% He an internal pressure level of $\sim$0.25~Pa is estimated after a period of 300 days, implying that under such conditions the SuperNEMO PMT will last 4-5 years before becoming difficult to operate. In practice it is expected that He concentrations surrounding SuperNEMO PMTs will be far lower than 10\% through effective gas-sealing of the tracking detector and flushing of the air volumes around the optical modules. The techniques presented in this article will provide a robust in-situ method of monitoring the He exposure of the PMTs throughout the lifetime of the experiment, which has now completed its initial commissioning phase~\cite{SuperNEMO:2024nqz}.
    
\acknowledgments
    We thank Hamamatsu Photonics for the loan of PMTs used in this experiment and for providing information on the glass composition. We acknowledge support by the MEYS of the Czech Republic (Contract Number LM2023063), CNRS/IN2P3 in France, APVV in Slovakia (Projects No. 15-0576 and 21-0377), NRFU in Ukraine (Grant No. 2023.03/0213), STFC in the UK, and NSF in the USA.

\bibliographystyle{rsc}
\bibliography{mybibfile}

\providecommand*{\mcitethebibliography}{\thebibliography}
\csname @ifundefined\endcsname{endmcitethebibliography}
{\let\endmcitethebibliography\endthebibliography}{}
\begin{mcitethebibliography}{16}
\providecommand*{\natexlab}[1]{#1}
\providecommand*{\mciteSetBstSublistMode}[1]{}
\providecommand*{\mciteSetBstMaxWidthForm}[2]{}
\providecommand*{\mciteBstWouldAddEndPuncttrue}
  {\def\EndOfBibitem{\unskip.}}
\providecommand*{\mciteBstWouldAddEndPunctfalse}
  {\let\EndOfBibitem\relax}
\providecommand*{\mciteSetBstMidEndSepPunct}[3]{}
\providecommand*{\mciteSetBstSublistLabelBeginEnd}[3]{}
\providecommand*{\EndOfBibitem}{}
\mciteSetBstSublistMode{f}
\mciteSetBstMaxWidthForm{subitem}
{(\emph{\alph{mcitesubitemcount}})}
\mciteSetBstSublistLabelBeginEnd{\mcitemaxwidthsubitemform\space}
{\relax}{\relax}

\bibitem[Incandela \emph{et~al.}(1988)Incandela, Ahlen, Beatty, Ciocio, Felcini, Ficenec, Hazen, Levin, Marin, Stone, Sulak, and Worstell]{Incandela:1987dh}
J.~Incandela, S.~Ahlen, J.~Beatty, A.~Ciocio, M.~Felcini, D.~Ficenec, E.~Hazen, D.~Levin, A.~Marin, J.~Stone, L.~Sulak and W.~Worstell, \emph{Nuclear Instruments and Methods in Physics Research Section A: Accelerators, Spectrometers, Detectors and Associated Equipment}, 1988, \textbf{269}, 237--245\relax
\mciteBstWouldAddEndPuncttrue
\mciteSetBstMidEndSepPunct{\mcitedefaultmidpunct}
{\mcitedefaultendpunct}{\mcitedefaultseppunct}\relax
\EndOfBibitem
\bibitem[Ospanov \emph{et~al.}(2019)Ospanov, Kordosky, Lang, Liu, Osiecki, Proga, and Vahle]{ospanov2019studiesheliumpoisoninghamamatsu}
R.~Ospanov, M.~Kordosky, K.~Lang, J.~Liu, T.~Osiecki, M.~Proga and P.~Vahle, \emph{Studies of helium poisoning of a Hamamatsu R5900-00-M16 photomultiplier}, 2019, \url{https://arxiv.org/abs/1908.08869}\relax
\mciteBstWouldAddEndPuncttrue
\mciteSetBstMidEndSepPunct{\mcitedefaultmidpunct}
{\mcitedefaultendpunct}{\mcitedefaultseppunct}\relax
\EndOfBibitem
\bibitem[Ma \emph{et~al.}(2011)Ma\emph{et~al.}]{Ma:2009aw}
K.~Ma \emph{et~al.}, \emph{Nuclear Instruments and Methods in Physics Research Section A: Accelerators, Spectrometers, Detectors and Associated Equipment}, 2011, \textbf{629}, 93--100\relax
\mciteBstWouldAddEndPuncttrue
\mciteSetBstMidEndSepPunct{\mcitedefaultmidpunct}
{\mcitedefaultendpunct}{\mcitedefaultseppunct}\relax
\EndOfBibitem
\bibitem[Arnold \emph{et~al.}(2010)Arnold\emph{et~al.}]{SuperNEMO:2010wnd}
R.~Arnold \emph{et~al.}, \emph{The European Physical Journal C}, 2010, \textbf{70}, 927--943\relax
\mciteBstWouldAddEndPuncttrue
\mciteSetBstMidEndSepPunct{\mcitedefaultmidpunct}
{\mcitedefaultendpunct}{\mcitedefaultseppunct}\relax
\EndOfBibitem
\bibitem[Barabash \emph{et~al.}(2017)Barabash\emph{et~al.}]{Barabash:2017sxf}
A.~Barabash \emph{et~al.}, \emph{Nuclear Instruments and Methods in Physics Research Section A: Accelerators, Spectrometers, Detectors and Associated Equipment}, 2017, \textbf{868}, 98--108\relax
\mciteBstWouldAddEndPuncttrue
\mciteSetBstMidEndSepPunct{\mcitedefaultmidpunct}
{\mcitedefaultendpunct}{\mcitedefaultseppunct}\relax
\EndOfBibitem
\bibitem[Townsend(1910)]{Townsend}
J.~Townsend, \emph{{The Theory of Ionization of Gases by Collision }}, London :Constable, 1910\relax
\mciteBstWouldAddEndPuncttrue
\mciteSetBstMidEndSepPunct{\mcitedefaultmidpunct}
{\mcitedefaultendpunct}{\mcitedefaultseppunct}\relax
\EndOfBibitem
\bibitem[Shah \emph{et~al.}(1988)Shah, Elliott, McCallion, and Gilbody]{Shah_1988}
M.~B. Shah, D.~S. Elliott, P.~McCallion and H.~B. Gilbody, \emph{Journal of Physics B: Atomic, Molecular and Optical Physics}, 1988, \textbf{21}, 2751--2761\relax
\mciteBstWouldAddEndPuncttrue
\mciteSetBstMidEndSepPunct{\mcitedefaultmidpunct}
{\mcitedefaultendpunct}{\mcitedefaultseppunct}\relax
\EndOfBibitem
\bibitem[ET (Accessed on 14/01/2024 \url{https://et-enterprises.com/images/brochures/Understanding\_Pmts.pdf})]{ETEnterprises}
ET Enterprises Online, ref:upmt/11, \emph{Understanding Photomultipliers}, Accessed on 14/01/2024 \url{https://et-enterprises.com/images/brochures/Understanding\_Pmts.pdf}\relax
\mciteBstWouldAddEndPuncttrue
\mciteSetBstMidEndSepPunct{\mcitedefaultmidpunct}
{\mcitedefaultendpunct}{\mcitedefaultseppunct}\relax
\EndOfBibitem
\bibitem[Altemose(1961)]{Altemose}
V.~O. Altemose, \emph{Journal of Applied Physics}, 1961, \textbf{32}, 1309--1316\relax
\mciteBstWouldAddEndPuncttrue
\mciteSetBstMidEndSepPunct{\mcitedefaultmidpunct}
{\mcitedefaultendpunct}{\mcitedefaultseppunct}\relax
\EndOfBibitem
\bibitem[Barrer(1941)]{Fick}
R.~M. Barrer, \emph{Diffusion in and through solids}, Cambridge University Press, New York, 1941\relax
\mciteBstWouldAddEndPuncttrue
\mciteSetBstMidEndSepPunct{\mcitedefaultmidpunct}
{\mcitedefaultendpunct}{\mcitedefaultseppunct}\relax
\EndOfBibitem
\bibitem[Quinn(2023)]{wq_thesis}
W.~S. Quinn, \emph{PhD thesis}, University College London, United Kingdom, 2023\relax
\mciteBstWouldAddEndPuncttrue
\mciteSetBstMidEndSepPunct{\mcitedefaultmidpunct}
{\mcitedefaultendpunct}{\mcitedefaultseppunct}\relax
\EndOfBibitem
\bibitem[CAE(Accessed on 14/01/2024 \url{https://www.caen.it/products/dt5751/})]{CAEN}
CAEN Online, \emph{DT5751 2/4 Channel 10 bit 2/1 GS/s Digitizer}, Accessed on 14/01/2024 \url{https://www.caen.it/products/dt5751/}\relax
\mciteBstWouldAddEndPuncttrue
\mciteSetBstMidEndSepPunct{\mcitedefaultmidpunct}
{\mcitedefaultendpunct}{\mcitedefaultseppunct}\relax
\EndOfBibitem
\bibitem[Huang(2005)]{HUANG2005811}
Y.-F. Huang, in \emph{The Electrical Engineering Handbook}, ed. W.-K. CHEN, Academic Press, Burlington, 2005, p. 811\relax
\mciteBstWouldAddEndPuncttrue
\mciteSetBstMidEndSepPunct{\mcitedefaultmidpunct}
{\mcitedefaultendpunct}{\mcitedefaultseppunct}\relax
\EndOfBibitem
\bibitem[Kondev and Lalkovski(2011)]{KONDEV2011707}
F.~Kondev and S.~Lalkovski, \emph{Nuclear Data Sheets}, 2011, \textbf{112}, 707--853\relax
\mciteBstWouldAddEndPuncttrue
\mciteSetBstMidEndSepPunct{\mcitedefaultmidpunct}
{\mcitedefaultendpunct}{\mcitedefaultseppunct}\relax
\EndOfBibitem
\bibitem[Acc()]{Oxides}
According to private and confidential correspondence with Hamamatsu Photonics for the percentage of glass forming oxides, \textit{M}, for the borosilicate glass used in the SuperNEMO R5912-MOD PMTs.\relax
\mciteBstWouldAddEndPunctfalse
\mciteSetBstMidEndSepPunct{\mcitedefaultmidpunct}
{}{\mcitedefaultseppunct}\relax
\EndOfBibitem
\bibitem[Aguerre \emph{et~al.}(2024)Aguerre\emph{et~al.}]{SuperNEMO:2024nqz}
X.~Aguerre \emph{et~al.}, \emph{{Commissioning of the calorimeter of the SuperNEMO demonstrator}}, 2024, \url{https://arxiv.org/pdf/2412.18021}\relax
\mciteBstWouldAddEndPuncttrue
\mciteSetBstMidEndSepPunct{\mcitedefaultmidpunct}
{\mcitedefaultendpunct}{\mcitedefaultseppunct}\relax
\EndOfBibitem
\end{mcitethebibliography}

\end{document}